\documentclass[aps,prb,floatfix,twocolumn,footinbib]{revtex4-1} 
\usepackage{epsfig}
\usepackage{hyperref}
\usepackage{cancel} 
\usepackage{amsmath, amssymb} 
\usepackage{graphicx}
\usepackage{epsf}
\usepackage{braket}
\usepackage{appendix}
\usepackage{color}
\usepackage{gensymb}

\begin{document}

\title{Interplay between anisotropy and spatial dispersion in metamaterial waveguide}

\author{K.\,L. Koshelev$^{1,2}$}
\email{ki.koshelev@gmail.com}
\author{A.\,A. Bogdanov$^{1,2}$}
\email{bogdanov@ioffe.mail.ru}

\affiliation{
$^{1}$ITMO University, 197101 St.~Petersburg, Russian Federation \\
$^{2}$Ioffe Institute, 194021 St.~Petersburg, Russian Federation}

\date{\today}

\begin{abstract}

We analyze spectrum of waveguide modes of an arbitrary uniaxial anisotropic metamaterial slab with non-local electromagnetic response whose permittivity tensor could be described within Drude approximation. Spatial dispersion was introduced within the hydrodynamical model. Both  anisotropy and spatial dispersion were considered as perturbations. This helps to distinguish their effect on the spectrum of the slab and to analyze lifting of the degeneracy of eigenmodes at plasma frequency in detail. Spatial dispersion is shown to result in break of the singularity in the density of optical states in the hyperbolic regime and in suppression of negative dispersion induced by anisotropy.  Mutual effect of spatial dispersion and anisotropy can bring light  to a complete stop at certain frequencies.

\end{abstract}


\maketitle

\section{Introduction}

Electromagnetic response of metamaterials in the simplest case is described by an effective permittivity and permeability, $\varepsilon$ and $\mu$. Spatial inhomogeneity and retardation effect result in dependance of the effective parameters on the frequency $\omega$ and the wavevector $\mathbf{k}$ of the incident wave. Anisotropic, chiral, and bianisotropic metamaterials are described by tensorial effective parameters $\hat\varepsilon(\omega,\mathbf{k})$ and $\hat\mu(\omega,\mathbf{k})$. The specific form of $\hat\varepsilon(\omega,\mathbf{k})$ and $\hat\mu(\omega,\mathbf{k})$ depends on the design of metamaterials. However, in the case of the long wavelength limit ($|\mathbf{k}|L\ll1$, where $L$ is character period of the structure) a local electromagnetic response can be often described in the framework of the Drude approach:\cite{Brown1953,Pendry1996,Pokrovsky2002,Silveirinha2005,Hoffman2007,Simovski2012, Koshelev2015,Bogdanov2011,Chebykin2011,Silveirinha2009,Chern2013}  
\begin{equation}
\label{DrudeLorentz}
\varepsilon(\omega)=\varepsilon_{\infty}\left(1-\frac{\Omega^{2}}{\omega(\omega+i\gamma)}\right).
\end{equation}
Here,  $\varepsilon_{\infty}$ is the permittivity of a host material, $\gamma$ is the damping parameter, $\Omega$ is the resonance frequency (plasma frequency) of a metamaterial. In isotropic metamaterials,  the resonance frequency $\Omega$ is degenerated.~\cite{Silveirinha2005,Silveirinha2009a} Structural anisotropy, i.e. anisotropy of meta-atoms or lattice of metamaterial can lift the degeneracy and dramatically change its properties. For example, anisotropy of effective masses of charge carriers in conducting layers of periodic metal-dielectric structures results in appearance of  additional allowed energy bands for photons.\cite{Bogdanov2012} Hyperbolic regime of metamaterial characterized by a singular density of optical states can be reached in media with anisotropic plasma frequency.\cite{Koshelev2015}  The structural anisotropy can be simply tailored at the fabrication stage.

Along with a structural anisotropy it is possible to distinguish anisotropy induced by spatial dispersion when an incident electromagnetic wave creates a preferential direction parallel to the wavevector $\mathbf{k}$ which plays a role of an optical axis.\cite{ginzburg1970, Chebykin2015}  Usually, in natural media, spatial dispersion is essential only in the vicinity of resonances (interband transitions, exciton absorbption, plasmon excitation etc) and can be neglected far from them.\cite{Pekar1958,Rao1978} In contrast to that, spatial dispersion in artificial media can be essential even in the long wavelength limit.\cite{Belov2003}

In the present paper we analyse and compare the effects of the spatial dispersion and structural anisotropy on the spectrum of a metamaterial slab. Dielectric function of the slab we describe within the Drude approximation. It is quite general approach since many types metamaterials from split-ring resonator based structures to wire or multilayer media can be described within it.\cite{Chebykin2011,Silveirinha2009,Koshelev2015,Simovski2012,Silveirinha2005} Anisotropy of the slab is introduced through the anisotropy of the plasma frequency. Spatial dispersion is considered within hydrodynamical approximation. In order to distinguish the effects of spatial dispersion and structural anisotropy we consider them as perturbations.  



The paper is organized as follows.  In Sec.~\ref{sec2a} we briefly discuss spectrum of bulk isotropic metamaterial. In Secs.~\ref{sec2b} and \ref{sec2c}, we consistently analyze spectra of isotropic and anisotropic metamaterial slab neglecting any spatial dispersion effects. In Secs.~\ref{sec2d} and \ref{sec2e}, we study the effects of non-local electromagnetic response on waveguide spectrum of isotropic and anisotropic metamaterial slab, respectively. Section~\ref{sec3} contains an analysis of a finite dielectric contrast between the cladding layers and the slab. In Sec.~\ref{sec4} we discuss a figure of merit and dissipation spectra. Finally, in Sec.\ref{sec5} we summarize our major results.



\section{Guided modes dispersion}


\subsection{Bulk metamaterial}
\label{sec2a}

Before turning to the problem of a metamaterial slab, let us briefly consider eigenmode spectrum of a bulk isotropic metamaterial with an arbitrary scalar permittivity $\varepsilon(\omega,\mathbf{k})$.
Spatial and time Fourier transform of Maxwell's equation $\nabla\cdot \mathbf{D}=0$ in the isotropic case yields:
\begin{equation}
\label{plasma}
\varepsilon(\omega,\mathbf{k})[\mathbf{k} \cdot \mathbf{E}(\omega,\mathbf{k}) ]=0
\end{equation}
Equation~(\ref{plasma}) has two solutions: (i) the first one is transversal electromagnetic waves satisfying the condition $ \mathbf{E} \bot  \mathbf{k}$; (ii) the second one is longitudinal waves satisfying the equation $\varepsilon(\omega,\mathbf{k})=0$.\footnote{Under the term \textit{longitudinal mode} we understand the mode for which wavevector $\mathbf{k}$ is collinear to the electric field $\mathbf{E}$.}  The longitudinal waves are pure electric ($\mathbf{H} =0$).\cite{ginzburg1970} In a plasma, metal or semiconductor, they represent oscillations of charge carrier density and often called bulk plasma waves or Langmuir waves~\cite{Tonks1929, shimotsuma2003self}. In a medium with local electromagnetic response, frequency of the Langmuir waves does not depend on both direction and absolute value of wavevector $\mathbf{k}$. Thus, the Langmuir waves form an infinite set of degenerated modes with zero group velocity.



Spatial dispersion results in a dependence of the longitudinal wave's frequency on absolute value of $\mathbf{k}$ but not on its direction. So, the degeneracy is lifted partly. Total lift of the degeneracy demands existence of an additional preferential direction not parallel  to $\mathbf{k}$. In a bulk medium, it can be induced, for example, by an external magnetic field  or by anisotropy of effective mass of carriers.\cite{Ivanov1999,Bogdanov2011} In the case of a slab, the preferential direction is naturally determined by the normal to the slab's interfaces.



\subsection{Isotropic slab}
\label{sec2b}

\begin{figure}[t]
\center{\includegraphics[width=0.7\linewidth]{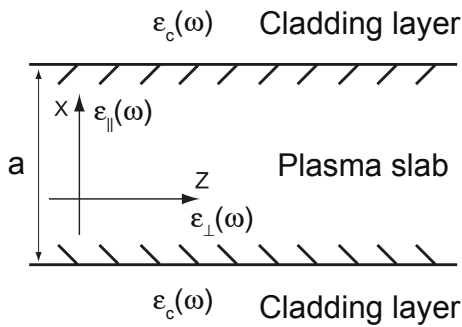}}
\caption{Metamaterial slab with plasma core and arbitrary cladding layers. Guided modes propagate along z-axis.}
\label{fig:waveguide}
\end{figure}
Let us consider a metamaterial slab with local isotropic electromagnetic response described in the framework of the Drude approach [see Eq.~(\ref{DrudeLorentz})]. Spectrum of such a slab consists of three types of eigenmodes: (i)~bulk waveguide modes formed due to total reflection of electromagnetic waves from the slab boundaries, (ii) two surface modes formed due to the constructive and destructive interference of the surface waves localized at the slab's boundaries, (iii) Langmuir modes formed due to reflection of pure electric longitudinal waves from the slab's boundaries. The properties of the bulk and surface modes are well-documented (see, e.g., Refs.~[\onlinecite{Doerr2008,Collin1991,Barnes2003,Pitarke2007,Polo2011}]), so, we manly focus on the properties of the Langmuir ones. The latter can be TM-polarization only and, therefore, have two components of electric field satisfying the equation:  
\begin{equation}
E_x=-\frac{i}{k_z}\frac{\partial E_z}{\partial x}.
\end{equation}
Here $k_z$ is the lateral component of wave vector. As well as in unbound metamaterial, the Langmuir modes do not satisfy the Helmholtz equation being at the same time a solution of the Maxwell's equation $\text{div}\mathbf{D}=0$.
Straightforward analysis of the Maxwell's equations and matching conditions at the boundaries shows that electric field of the Langmuir modes is completely confined in the slab and does not penetrate into the cladding layers independently on their permittivity. The frequency of the Langmuir modes does not depend on $k_z$ and coincides with plasma frequency of the slab $\Omega$ as in a bulk metamaterial. Therefore, in an isotropic metamaterial slab Langmuir waves represent a degenerated set of of eigenmodes with zero group velocity. Introducing the losses ($\gamma\neq0$) into the slab does lift the degeneracy making the frequency a complex value $\omega\approx\Omega+i\gamma/2$.

An example of the mode structure of an isotropic metamaterial slab is shown in Fig.~\ref{fig:dispersion}(a). For the sake of simplicity, the cladding layers is assumed to be a perfect conductor. Effect of a finite dielectric contrast between the cladding layers and the slab is considered further in Sec.~\ref{sec:anis_spat}.   



\subsection{Anisotropic slab}
\label{sec2c}

Let us consider a metamaterial slab with uniaxial anisotropic permittivity with the optical axis parallel to the $x$-direction. The permittivity tensor in this case is given by 
\begin{equation}
\label{eps_tensor}
   \hat \varepsilon_{\rm{s}} (\omega)=
   \left(
   \begin{matrix} 
      \varepsilon_{\|}(\omega) & 0 & 0\\
      0 & \varepsilon_{\bot}(\omega) & 0\\
      0 & 0 & \varepsilon_{\bot}(\omega) \\
   \end{matrix}
   \right).
\end{equation}
Here,  indices ${\|}$ and $\bot$ corresponds to direction along and across the optical axis. We suppose that the tensor components have Drude dispersion [see Eq.~(\ref{DrudeLorentz})] with different plasma frequencies $\Omega_\bot$ and $\Omega_\|$. Host permittivity $\varepsilon_\infty$ is supposed to be isotropic.
 
The Helmholtz equation for the TM-polarized modes inside the slab is reduced to the following:
\begin{equation}
\label{anis_Helm}
\varepsilon_{\|}\frac{\partial^2 E_x}{\partial x^2}+\varepsilon_{\bot}\left(\varepsilon_{\|}\frac{\omega^2}{c^2}-k_z^2\right)E_x=0.
\end{equation}
Analytical expression for the dispersion of the Langmuir and bulk waveguide modes can be straightforwardly obtained from Eq.~(\ref{anis_Helm}) in the case of perfect electric conductor boundary condition:
\begin{equation}
\label{disp_an}
k_z^2=\varepsilon_\|\left(\frac{\omega^2}{c^2}-\frac{\pi^2n^2}{a^2\varepsilon_\bot}\right).
\end{equation}
Here, $n=0,1,2,...$ is an integer mode number. Figure~\ref{fig:dispersion}(b) shows the dispersion of Langmuir and bulk waveguide modes for the case $\Omega_\bot>\Omega_\|$. One can see that degeneracy for the Langmuir modes is lifted. Their spectrum sandwiched between plasma frequencies $\Omega_\bot$ and $\Omega_\|$, where $\varepsilon_\|\varepsilon_\bot<0$ and the metamaterial slab exhibits properties of a hyperbolic medium.\cite{Poddubny2013} Density of states for the Langmuir modes is singular for any frequency $\Omega_\|<\omega<\Omega_\bot$ as all of the modes (except one corresponding to $\varepsilon_\bot(\omega)=0$) has the common frequency cutoff $\Omega_\|$ and the common horizontal asymptote $\omega=\Omega_\bot$. Only the fundamental Langmuir mode ($n=0$) is pure electrical. The rest Langmuir modes have non-zero magnetic field and, therefore, non-zero Pointing vector. Their group velocity $\mathbf{v}_\text{g}=\partial \omega /\partial k_z$ can be found from Eq.~(\ref{disp_an}). In the case of $\Omega_\|>\Omega_\bot$ dispersion of the Langmuir modes is negative  [see Fig.~\ref{fig:dispersion}(c)].         

In spite of existence of the analytical expression for the dispersion [see Eq.~(\ref{disp_an})], a more simple way to analyse spectrum of the Langmuir modes and to gain a deep insight into how the anisotropy lifts the degeneracy is to consider the anisotropy as a perturbation. Let us take the difference  between plasma frequencies along $\|$ and $\bot$ directions as a perturbation parameter ($\delta \Omega=\Omega_\bot-\Omega_\|$) assuming that $|\delta\Omega|~\ll~\Omega_\|$.

Using the perturbation theory for the case of degenerate spectrum one can find the set of zeroth-order eigenfuctions whose change under the action of the small applied perturbation is small:
\begin{equation}
\label{E_field}
E_z=E_0\sin\left(\frac{\pi n x}{a}\right), \ \ \ \  E_x=-E_0\frac{i \pi n}{a k_z}\cos\left(\frac{\pi n x}{a}\right).
\end{equation}
Here, $n=0,1,2,...$ is the mode number as in Eq.~(\ref{disp_an}).  It follows from Eq.~(\ref{E_field}) that the Langmuir modes are nearly longitudinal if $n\ll k_za$ and nearly transversal if $n\gg k_za$. 

Perturbation to the eigenfrequency of Langmiur mode $\delta \omega_n$ is readily found as:
\begin{equation}
\label{om_perturbation}
\delta\omega_n=\delta\Omega\frac{k_z^2}{k_z^2+\left(\frac{\pi n}{a}\right)^2}.
\end{equation}
On can see that perturbation theory works for all mode numbers $n$ for both short and long wavelength limits as $|\delta\omega_n|\leq|\delta\Omega|$ for all $k_z$. Equation (\ref{om_perturbation}) predicts right sign of group velocity for the Langmuir modes.    


As was mentioned earlier, in the case of anisotropic slab, the Langmuir modes are not pure electric waves. Their magnetic field is given by:
\begin{equation}
H_y=-2i\delta\Omega\varepsilon_\infty E_0\frac{\left(\frac{\pi n}{a}\right)}{c\left[k_z^2+\left(\frac{\pi n}{a}\right)^2\right]}\cos\left(\frac{\pi n x}{a}\right).
\end{equation}
One can see that magnetic field does not vanishes at the slab's boundaries and, therefore, penetrates inside the cladding layers. 

\subsection{Isotropic slab with effects of nonlocality}
\label{sec2d}

The problem of spatial dispersion has a long history and is still being discussed.
 \cite{Pekar1958,Leontovich1961,Sauter1967,agranovich2013crystal,Rukhadze1961, Agranovich1973,Forstmann1979,Forstmann1986,Davis1988,Scandolo2001,Belov2003} The main stumbling block of the spatial dispersion problem is additional boundary conditions. Their introduction is necessary because a nonlocal response increases the order of the Maxwell's equations but their choice is ambiguous.

In the presence of a weak nonlocality, electric induction $\mathbf{D}$ can be expressed through electric field $\mathbf{E}$  as (see, e.g., Ref.~[\onlinecite{agranovich2013crystal}])
\begin{equation}
\label{cc}
\mathbf{D}=\varepsilon_{\rm s}(\omega)\mathbf{E}+ C_1\text{grad}(\text{div}\mathbf{E}) + C_2\text{rot}(\text{rot}\mathbf{E}).
\end{equation}
The second term affects only the transversal waves while the first one only the longitudinal waves. The coefficients $C_1$ and $C_2$ can have a frequency dependance, which is determined by the particular physical model.  We will analyze the spatial dispersion within the hydrodynamical approximation which takes into account only the fist term of Eq.~(\ref{cc}). This approximation is often used for the description of plasma oscillations in condensed matter systems including wire media and nanoparticle composites.\cite{Barton1979,Boardman1982,Mortensen2013,Toscano2012,Christensen2014} It describes motion of the charges by the Euler equation thats results in the following dependance of $\mathbf{D}$ on $\mathbf{E}$:
\begin{equation}
\label{electrical_induction}
\mathbf{D}=\hat\varepsilon_{\rm s}\mathbf{E} + A\frac{v^2}{\omega(\omega+i\gamma)}\text{grad}(\text{div}\mathbf{E})
\end{equation}
Here, $A$ is a numerical constant and $v$ is the mean velocity of chaotic motion of charges. In the case of non-degenerated plasma  $v$ represents the thermal velocity of carriers and $A=1/3$.\cite{Forstmann1986} In the case of degenerate plasma $v$ is the Fermi velocity and $A=3/5$.\cite{Forstmann1986} Here we put $\gamma$=0. The effect of losses will be considered further in Sec.~\ref{sec:losses}.

Let us note that as follows from Eq.~(\ref{electrical_induction}) the nonlocal response is of the order of $(vk/\omega)^2$. Therefore, it is significant only for the modes with low phase velocity $\omega/k$ comparable with characteristic velocity of carriers $v$, i.e. only for the Langmuir and surface modes.


Substitution of electrical displacement $\mathbf{D}$ from Eq.~(\ref{electrical_induction}) into the Maxwell's equations yields the Helmholtz equation of the forth order for TM-polarized waves:
\begin{align}
\label{eq_spat_disp}
\left[\alpha\frac{\partial^2}{\partial x^2}
\!+\! \left(k_x^\text{s}\right)^2\right]\!\!\left[\frac{\partial^2}{\partial x^2}\!+\!\left(k_x^\text{f}\right)^2\!\right]\!E_x\!=\!0. \\
\left(k_x^\text{s}\right)^2=\frac{\varepsilon_{\rm s}\omega^2}{c^2} - \alpha k_z^2 \ \ \ \ \ \ \ \ \ \ \ \ \ \ \ \ \ \ \ \ \\
\left(k_x^\text{f}\right)^2=\frac{\varepsilon_s\omega^2}{c^2}-k_z^2 \ \ \ \ \ \ \ \ \ \ \ \ \ \ \ \ \ \ \ \ \ \,
\end{align}
%
The wavevector components $k_x^\text{s}$ and $k_x^\text{f}$ correspond to the fast (waveguide) and slow (Langmuir) modes, respectively. The parameter $\alpha=Av^2/c^2$ is the dimensionless quantity characterizing the spatial dispersion.

In virtue of the symmetry, the general solution inside the slab can be divided into symmetric and antisymmetric:  
\begin{equation}
E_x(x)=A   \left\{\begin{array}{cc} 
      \cos{(k_x^{\rm f} x)} \\
      \sin{(k_x^{\rm f} x)} \\
   \end{array}\right\}+B\left\{\begin{array}{cc} 
      \cos{(k_x^{\rm s} x)} \\
      \sin{(k_x^{\rm s} x)} \\
   \end{array}\right\}
\end{equation}
The magnetic field is non-zero only for the fast modes:
\begin{equation}
H_{y}(x)=A\frac{\varepsilon_{\rm s}\omega}{ck_z}
\left\{\begin{array}{cc} 
      \sin{(k_x^{\rm f} x)} \\
      \cos{(k_x^{\rm f} x)} \\
   \end{array}\right\}
\end{equation}
In the general case, eigenmodes represent a superposition of the fast and slow modes. The ratio between amplitudes $A$ and $B$ is determined from the additional boundary conditions. They depends on the model used to describe the spatial dispersion and, generally speaking, it should be determined microscopically.\cite{Agranovich1973,Maslovski2000} Here, we do not specify additional boundary conditions assuming the propagation of the fast and slow modes independent. Therefore, if penetration of the modes into the cladding layers is weak, the wavevectors $k_x^\text{f}$ and $k_x^\text{s}$ can be quantized independently as $\pi n/a$. Dispersions of the waveguide  and Langmuir modes are are given by:
\begin{figure*}[t]
\begin{minipage}{0.93\linewidth}
\center{\includegraphics[width=1\linewidth]{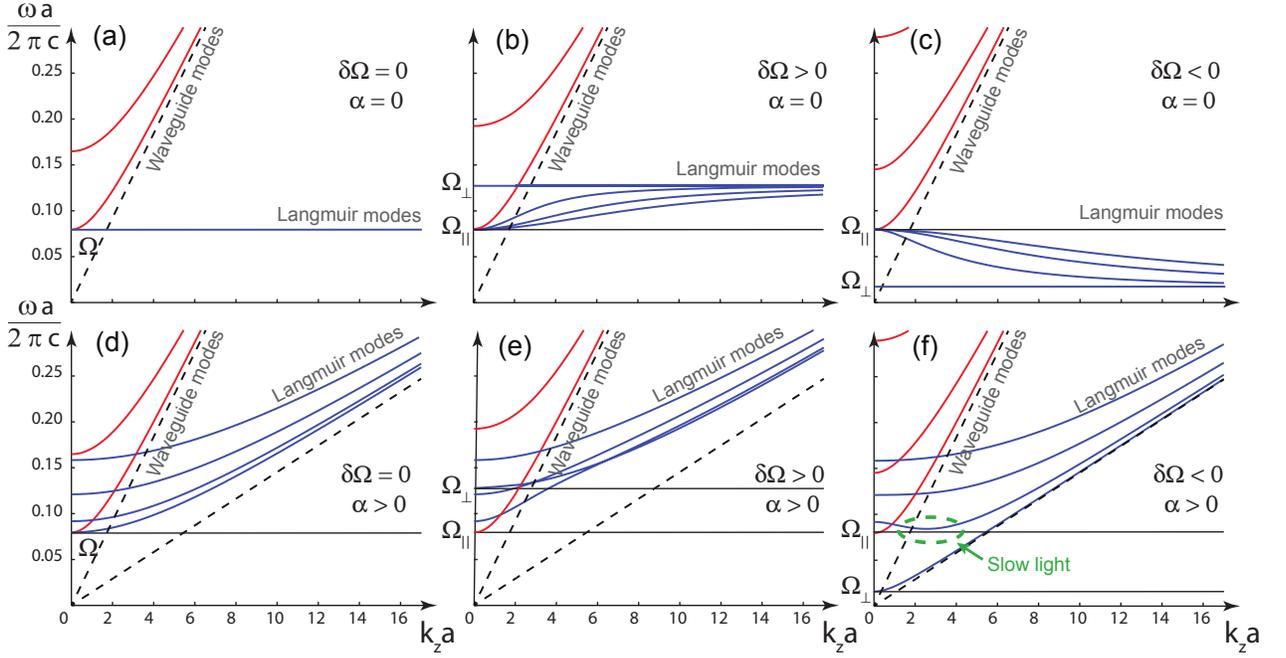}}
\caption{(Color online). Dispersion of TM-polarized guided modes propagating inside the metamaterial slab with metal claddings. Red solid lines show the dispersion of fast conventional waveguide modes, blue solid lines picture the dispersion of slow Langmuir modes. Black dashed lines represent the light lines $\omega=ck_z/\sqrt{\varepsilon_\infty}$ and $\omega=Avk_z/\sqrt{\varepsilon_\infty}$ which describe the asymptotical behaviour of fast and slow modes correspondingly. Black horizontal solid lines show plasma frequencies $\Omega_\|$ and $\Omega_\bot$.  Insets describe the regime, i.e. the magnitude of anisotropy ($\delta\Omega$) and non-local effects ($\alpha$). Parameters are $\alpha=0.1$, $\varepsilon_{\rm s}=12$.}
\label{fig:dispersion}
\end{minipage}
\end{figure*}
\begin{eqnarray}
\left(\omega_n^{\rm f}\right)^2&=&\Omega^2+\frac{c^2}{\varepsilon_{\infty}}\,\left[k_z^2+\left(\frac{\pi n}{a}\right)^2\right] \\
\left(\omega_n^{\rm s}\right)^2&=&\Omega^2+\frac{Av^2}{\varepsilon_{\infty}}\!\left[k_z^2+\left(\frac{\pi n}{a}\right)^2\right].
\end{eqnarray}
Their plot is shown in Fig.~\ref{fig:dispersion}(d) by red and blue lines, respectively. 
One can see that degeneracy of Langmuir modes is lifted. Moreover, spatial dispersion results in dependance of their cut-off frequencies on the mode number $n$. So, singularity in density of optical states is destructed because there is only a finite number of the modes at any fixed frequency $\omega$.

Let us note that the Pointing vector $\mathbf{S}$ defined as $\frac{c}{8\pi}\text{Re}(\mathbf{E}\times\mathbf{H})$ is zero for the Langmuir modes is zero as they are pure electric modes. However, their group velocity is non-zero. There is no a contradiction because the Langmuir modes transfer the energy accumulated by election gas due to its compression and expansion.\cite{Forstmann1979} Fast waveguide modes transfer energy due to non-zero both electric and magnetic fields. Therefore, we have different mechanisms of the energy transfer for the Langmuir and waveguide modes.

As was mentioned above, we neglect an interaction between the Langmuir and waveguide modes and, therefore, we avoid consideration of additional boundary conditions. Accounting for an interaction between the Langmuir and waveguide modes results in appearance of anti-crossing between their dispersion curves.\cite{Tyshetskiy2014} In a symmetric waveguide, the interaction is possible only between modes of the same parity. Strength of the splitting depends on the additional boundary conditions.

\subsection{Anisotropic slab with effects of nonlocality \label{sec:anis_spat}}
\label{sec2e}
Now let us take into account both anisotropy of the slab and non-locality within hydrodynamical approach. The Helmholtz equation for TM-polarized wave in this case can be obtained straightforwardly from the Maxwell's equations:





\begin{widetext}
\begin{equation}
\label{fourth-eq-2}
\alpha\frac{\partial^2}{\partial x^2}\left[\frac{\partial^2 E_x}{\partial x^2}+\left(\frac{\varepsilon_{\bot}\omega^2}{c^2}-k_z^2\right)E_x\right]
+\left(\frac{\varepsilon_{\|}\omega^2}{c^2}-\alpha k_z^2\right)\frac{\partial^2 E_x}{\partial x^2}
+\left(\frac{\varepsilon_{\bot}\omega^2}{c^2}-\alpha k_z^2\right)\left(\frac{\varepsilon_{\|}\omega^2}{c^2}-k_z^2\right)E_x=0.
\end{equation}
\end{widetext}
Equation (\ref{fourth-eq-2}) can be written in the compact form using the substitution $E_x \propto e^{\pm ik_x x}$:


\begin{equation}
\label{dis_hard}
\frac{\omega^2}{(1-\alpha)c^2}=\frac{k_x^2}{\varepsilon_{\bot}(\omega,k)}+\frac{k_z^2}{\varepsilon_{\|}(\omega,k)}.
\end{equation}
Here, we use the following notations:
\begin{equation}
\label{dis_hard_1}
\varepsilon_{\bot,\|}(\omega,k)=\varepsilon_{\bot,\|}(\omega)-\alpha\frac{c^2(k_x^2+k_z^2)}{\omega^2}.
\end{equation}
Equation (\ref{dis_hard}) is biquadratic. Its solutions $\pm k_x^{\rm f}$ and $\pm k_x^{\rm s}$ correspond to the slow and fast  modes with symmetric and anti-symmetric field distribution. Dispersion of the eigenmodes depends on boundary conditions which define a quantization rule for $k_x^{\rm f}$ and $k_x^{\rm s}$. As in the previous section, we will quantize $k_x^{\rm f}$ and $k_x^{\rm s}$ independently as $\pi n/a$ assuming that interaction between the slow and fast modes is vanishingly small  and penetration depth of the modes in the slab claddings is negligible. The dispersions $\omega_n^{\rm f,s}(k_z)$ found from Eq.~(\ref{dis_hard}) for the cases $\Omega_\| < \Omega_\bot$ and $\Omega_\| > \Omega_\bot$  are shown in Figs.~\ref{fig:dispersion}(e) and \ref{fig:dispersion}(f), receptively. The analytical expressions for $\omega_n^{\rm f,s}$ are cumbersome and not convenient for analysis. The compact expressions revealed particular contributions of anisotropy and non-local effects can be obtained within perturbation theory assuming that $|\delta\Omega| \ll\Omega_{\|,\bot}$ and $v\ll c$:   
\begin{eqnarray}
\!\!\left(\omega_n^{\rm f}\right)^2\!\!=\!\Omega^2\!\!\left[\!1\!+\!\frac{2\delta\Omega}{\Omega}\frac{\frac{\pi^2 n^2}{a^2}}{\frac{\pi^2 n^2}{a^2}\!+\!k_z^2}\!\right]\!\!+\!\frac{c^2}{\varepsilon_{\infty}}\!\!\left[k_z^2\!+\!\frac{\pi^2n^2}{a^2}\right] \  \ \\
\label{thatone}
\!\left(\omega_n^{\rm s}\right)^2\!\!=\!\Omega^2\!\!\left[\!1\!+\!\frac{2\delta\Omega}{\Omega}\frac{k_z^2}{\frac{\pi^2 n^2}{a^2}\!+\!k_z^2}\!\right]\!\!+\!\frac{A v^2}{\varepsilon_{\infty}}\!\!\left[k_z^2\!+\!\frac{\pi^2 n^2}{a^2}\right].
\label{thisone}
\end{eqnarray}
One can see that in contrast to  anisotropy, spatial dispersion affects only the Langmuir modes. Simple analysis of Eq.~(\ref{thisone}) yields that Langmuir modes exhibit negative dispersion if $\delta\Omega<0$ for $k_z<k_z^*$, where

\begin{equation}
\label{inequality}
k_z^*=\frac{\pi n}{a}\left(\frac{n^*}{n}-1\right)^{1/2}.
\end{equation}
\begin{figure}[t]
\begin{minipage}{0.94\linewidth}
\center{\includegraphics[width=1\linewidth]{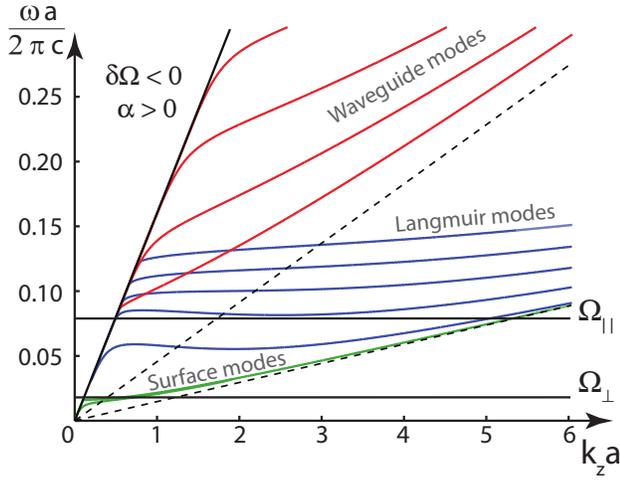}}
\caption{(Color online). Dispersion of TM-polarized guided modes propagating inside the metamaterial slab with dielectric claddings with constant permittivity $\varepsilon_{\rm c}$. Red solid lines show the dispersion of fast conventional waveguide modes, blue solid lines picture the dispersion of slow Langmuir modes, green solid lines show the dispersion of slow surface plasmon-polariton modes. Black dashed lines represent the light lines $\omega=ck_z/\sqrt{\varepsilon_\infty}$ and $\omega=Avk_z/\sqrt{\varepsilon_\infty}$ which describe the asymptotical behaviour of fast and slow modes correspondingly. Black inclined solid line represents the light line $\omega=ck_z/\sqrt{\varepsilon_{\rm c}}$ which separates the region of leaky and stable modes. Parameters are $\alpha=0.1$, $\varepsilon_{\rm s}=12$, $\varepsilon_{\rm c}=1$.}
\label{fig:dielectric}
\end{minipage}
\end{figure}
Here $n^*$ is maximal index of the Langmuir mode for which negative dispersion exists:
\begin{equation}
n^*=\frac{a}{\pi}\sqrt{\frac{2\Omega|\delta\Omega|\varepsilon_{\infty}}{A v^2}}.
\end{equation}
Spatial dispersion and anisotropy make oppositely directed contributions into the energy flow for the Langmuir modes. Therefore, the flows of electromagnetic and mechanic energy can completely compensate each other and bring the Langmuir mode to a complete stop. Slow light can be observed at the frequency $\omega_n^{\rm s}(k_z^*)$. This allows one to reach cavity regime without mirrors similar to distributed feedback cavities.\cite{Kogelnik1972, Kazarinov1973}

It should be mentioned that anisotropy does not affect the fundamental Langmuir mode ($n=0$), so, it remains pure electric ($H_y=0$) and longitudinal ($E_x=0$). Dispersion of the main Langmuir mode is determined by equation $\varepsilon_\bot(\omega,k_z)=0$. Therefore, it is always positive. For other Langmuir modes, the frequency bandwidth $\Delta\omega$ of negative dispersion depends on the mode number $n$ as:
\begin{equation}
\Delta\omega=\omega_n^{\rm s}(0)-\omega_n^{\rm s}(k_z^*)\approx|\delta\Omega|\left(\frac{n}{n^*}-1\right)^2.
\end{equation}



\section{Effect of cladding layers}
\label{sec3}

In the previous sections we neglect the penetration depth of the field inside the cladding layers assuming that dielectric contrast between them and the metamaterial slab is infinitely high. Within this assumption, $k_x^{\rm f}$ and $k_x^{\rm s}$ can be quantized independently as $\pi n/a$. Here, we analyze the effect of finite dielectric contrast on the dispersion of the eigenmodes.

A finite dielectric contrast results in appearance of two surface waves in the spectrum forming due to constructive and destructive interference of surface plasmon polariton (SPP) modes localized at slab's interfaces. Electromagnetic properties of SPPs are well-documented and we do not focus on them.\cite{agranovich2012surface,Barnes2003,Pitarke2007,Maier2005,Polo2011,Raza2013,Bogdanov2011}  

Penetration of the modes inside the cladding layers effectively increases the thickness of the slab. The correction to $k_x^{\rm f}$ in the isotropic case within the assumption of high dielectric contrast ($|\varepsilon_s/\varepsilon_c|\gg1$) can be derived from the expression for the Goos-Haanchen shift:\cite{Renard1964,Snyder1976,Mcguirk1977,Wild1982,Gehring2013}
\begin{equation}
\delta k_{x,n}^{ f} = - \frac{\varepsilon_c\varepsilon_s\omega^2/c^2}{\sqrt{k_z^2-\varepsilon_{\rm c}\omega^2/c^2}}\frac{2\pi n}{\varepsilon_{\rm s}(\omega)k_z^2a^2-\varepsilon_{\rm c}\pi^2n^2}
\end{equation}
Here $\varepsilon_c$ is the permittivity of the cladding layers. One can see that in the case of metal cladding layers $\varepsilon_c<0$ the correction is pure real. However, in the case of dielectric cladding layers $\varepsilon_c>0$, the correction is real under light line $\varepsilon_c^{1/2}\omega/c=k_z$ and imaginary above the light line where the modes are leaky. 

It should be mentioned that, in the isotropic case and when we neglect a mixing between the Langmuir and waveguide modes, the cladding layers do not affect the dispersion of the Langmuir modes at all since they are perfectly confined inside the slab. Therefore, in the framework of this approach, the Langmuir modes remain non-leaky above the light line of  the cladding layers in sharp contrast to the waveguide modes. However, more deep analysis shows that the Langmuir modes penetrate into the cladding layers and have leakage losses above the light line. It occues because of their mixing with waveguide modes which is determined by additionally boundary conditions.

\begin{figure}[t]
\begin{minipage}{1\linewidth}
\center{\includegraphics[width=1\linewidth]{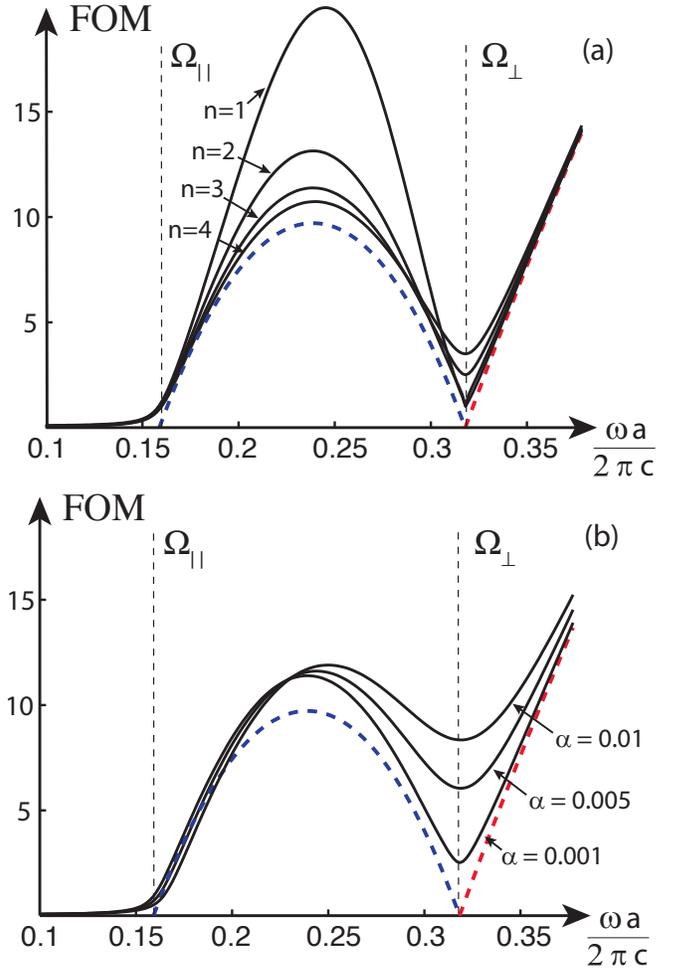}}
\caption{(Color online). Dependence of Figure of Merit $\eta$ on the frequency $\omega$ in the anisotropic metamaterial slab with non-local electromagnetic response. Blue and red dashed lines represent analytical expressions Eq.~\eqref{fom_1} and Eq.~\eqref{fom_2}, respectively. Parameters are $\gamma a/2\pi c=0.1$, $\varepsilon_{\rm s}=12$. (a)~Dependence of $\eta$ on $\omega$ for different mode numbers $n$. Parameter $\alpha$ is assumed to be equal to 0.001. (b)~Dependence of $\eta$ on $\omega$ for different $\alpha$. Mode number $n$ is assumed to be equal to $3$.}
\label{fig:fom}
\end{minipage}
\end{figure}

The dispersion $\omega(k_z)$ of TM-polarized modes inside the anisotropic plasma slab with dielectric claddings ($\varepsilon_c>0$) in the presence of spatial dispersion is shown in Fig.~\ref{fig:dielectric}. Parameters the structure are described in the capture of the figure. Fast, slow, and surface modes are shown by red, blue and green lines, respectevely.

\section{Losses\label{sec:losses}}
\label{sec4}

Dissipation spectra of the waveguide and surface modes are well-documented (see, e.g., Refs.,~[\onlinecite{adams1981introduction,Dionne2005,Bogdanov2012a}]), so, here, we focused on the Langmuir modes only. In the case of non-zero losses  [$\gamma\neq0$ in Eq.~\eqref{DrudeLorentz}] figure of merit (FOM) for the modes can be introduced as:
\begin{equation}
\eta(\omega)=\frac{{\rm Re}(k_z) }{{\rm Im}(k_z)}.
\end{equation}
The meaning of such defined $\eta(\omega)$ is the free path measured in the wavelengths. Analytical expression for $\eta(\omega)$ of the Langmuir modes in the hyperbolic regime neglecting the spatial dispersion can be carried out straightforwardly from  Eqs.~\eqref{dis_hard} and \eqref{dis_hard_1}:
\begin{equation}
\eta(\omega)=\frac{2}{\gamma\omega}\left|\frac{(\omega^2-\Omega_\bot^2)(\omega^2-\Omega_\|^2)}{\Omega_\|^2-\Omega_\bot^2}\right|.
\label{fom_1}
\end{equation}
One can see that FOM reaches maximum value
\begin{equation}
\eta_{\rm max}=\frac{|\Omega_\bot-\Omega_\| |}{\gamma} \ \ \ \text{at} \ \ \ \omega^*=\sqrt{\frac{\Omega^2_\|+\Omega^2_\bot}{2}}.
\end{equation}
Comparison of  $\eta(\omega)$ with results of numerical simulations [Figs.~\ref{fig:fom}(a) and \ref{fig:fom}(b)] shows that Eq.~\eqref{fom_1} works well for high mode numbers ($n\gg1$).

Propagation of the Langmuir modes at high frequencies ($\omega>\Omega_{\bot,\|}$) is possible only in a spatial dispersive media. In this case, the anisotropy is not essential ($\varepsilon_\bot\approx\varepsilon_\|$) and FOM can be estimated from  Eqs.~\eqref{dis_hard} and \eqref{dis_hard_1} as    
\begin{equation}
\eta(\omega)=\frac{2}{\gamma\omega}(\omega^2-\Omega_\bot^2).
\label{fom_2}
\end{equation}
Comparison of  this expression with results of numerical simulations [Figs.~\ref{fig:fom}(a) and \ref{fig:fom}(b)] shows that it works well for low mode numbers $n$.

\section{Summary \label{sec5}}
In this paper, we developed the theory of anisotropic metamaterial waveguide with non-local electromagnetic response. Anisotropy and spatial dispersion were taken into account as perturbation that allows one to distinguish their effect on the waveguide spectrum.

 It was shown that anisotropy of plasma oscillations lift the degeneracy of the Langmuir modes keeping their density of states singular. Whereas, even small spatial dispersion destroys the singularity. 

Spatial dispersion and anisotropy can make oppositely directed contributions into the energy flow for the Langmuir modes. Such interplay can bring light to a complete stop. This allows one to reach cavity regime without any mirrors similar to distributed feedback cavities.  

We have shown that the Langmuir modes in an isotropic waveguide are perfectly confined even above the light line of the cladding layers. They may become leaky either because of anisotropy of the waveguide or because of the resonance mixing with leaky modes, which occurs due to the spatial dispersion effects.

\acknowledgments

This work has been supported by RFBR (16-37-60064, 15-32-20665), by the President of Russian Federation (MK- 6462.2016.2), the Federal Programme on Support of Leading Scientific Schools (NSh-5062.2014.2), and by program of Fundamental Research in Nanotechnology and Nanomaterials of the Russian Academy of Science. Numerical simulations have been supported by the Russian Science Foundation (Grant \#15-12-20028).

\bibliography{references}

 \end{document}